# First principles design of divacancy defected graphene nanoribbon based rectifying and negative differential resistance device

Soubhik Chakrabarty,[a] A. H. M. Abdul Wasey,[a] Ranjit Thapa,[b,*] G. P. Das[a,*]

[a]Department of Materials Science, Indian Association for the Cultivation of Science, Jadavpur, Kolkata-700032, India

[b]SRM Research Institute, SRM University, Kattankulathur - 603203, Tamil Nadu, India.

*Corresponding Author E-mail: msgpd@iacs.res.in (GPD), ranjit.t@res.srmuniv.ac.in (RT)

**ABSTRACT**

We have elaborately studied the electronic structure of 555-777 divacancy (DV) defected armchair edged graphene nanoribbon (AGNR) and transport properties of AGNR based two-terminal device constructed with one defected electrode and one N doped electrode, by using density functional theory and non-equilibrium Green's function based approach. The introduction of 555-777 DV defect into AGNRs, results in a shifting of the $\pi$ and $\pi^*$ bands towards the higher energy value which indicates a shifting of the Fermi level towards the lower energy. Formation of a potential barrier, very similar to that of conventional *p-n* junction, has been observed across the junction of defected and N doped AGNR. The prominent asymmetric feature of the current in the positive and negative bias indicates the diode like property of the device with high rectifying efficiency within wide range of bias voltages. The device also shows robust negative differential resistance (NDR) with very high peak-to-valley ratio. The analysis of the shifting of the energy states of the electrodes and the modification of the transmission function with applied bias provides an insight into the nonlinearity and asymmetry observed in the I-V characteristics. Variation of the transport properties on the width of the ribbon has also been discussed.

# 1. Introduction

The remarkable technological progress and reduction in the size of electronic devices over the last two decades have driven the scientific community and industries to find the significance of nanomaterials in the electronic devices such as rectifiers,[1,2] switching devices,[3,4] field effect devices,[5,6] spin-filters,[7-9] negative differential resistance (NDR) based devices,[10-12] etc. Particularly at present rapid developments of NDR and rectifying devices are going on, which include nanoribbon,[12] nanowire,[13] quantum dot,[14] nanotube,[1,11] molecular junction[15] etc. Also the search of suitable nanostructures having tunable topology with versatile electronic properties is going on to look beyond Si based devices. In the recent past, graphene has attracted a lot of attention, due to its unique physical properties, high mobility, low power consumption and above all the ease to synthesize.[16-20] However the zero band gap semi-metallic nature of graphene is not suitable for device application. Tailoring the width of graphene sheet less than 10 nm results in one-dimensional (1D) graphene nanoribbon (GNR) that exhibits finite band-gap.[21] Depending on the edge geometry there are two primary types of GNR, *viz.* (i) Armchair edged graphene nanoribbon (AGNR) and (ii) Zigzag edged graphene nanoribbon (ZGNR). Both these GNR with each edge atoms passivated by single hydrogen atom, exhibit energy gaps that decrease with increasing GNR width.[22] Moreover, the electronic properties of GNR are very sensitive to many other factors, such as application of electric field,[23] modification of edges,[24-26] doping,[6,12,27-30] introduction of topological defects,[31-34] chemical functionalization[35,36] etc. Such wide range of functionalities[37] of GNR has established it as a potential candidate for the post-Si device applications.

Doping is one of the most fundamental and frequently used ways to tailor the electronic property of GNR. There are mainly two different processes of doping (i) doping with foreign elements and (ii) self doping by introduction of defects. In the context of foreign element doping, the introduction of Boron (B) and Nitrogen (N) resulting in hole and electron doped GNR, has been reported in the literature.[6,28,38,39] Several B and N doped GNR based rectifiers and NDR devices have been reported and underlying mechanisms have been proposed.[12,40-44] The modification of transmission function due to the shifting of electrode energy levels with applied bias was proposed to be the main reason behind the two above phenomena as reported by Zhang et al.[42] Pramanik et al.[40] explained the origin of rectification and NDR of a B and N doped AGNR based device on the basis of relative shifting of different energy levels of the total system with applied source-to-drain voltage. Deng et al.[41] explained the observed rectifying behavior of a zigzag-edged trigonal GNR device in terms of the asymmetric distribution of the electrostatic potential across the devices and the spatial distribution of electronic states at different applied voltages. Width dependent rectifying character was observed by Zheng et al.[44] in a Z-shaped GNR device which was explained by the analysis of spatial distribution of molecular energy levels. On the other hand self-doping[18,45] *via* defects interaction plays an important role in the modification of electronic structure of GNR due to disorder and localization. Topsakal et al.[46] have shown that vacancy defect in AGNR leads to a modification of band-gap that depends on the position of the defects with respect to the edges. Suppression of conductance was observed in vacancy defected GNR originating from the localization of electronic states that eventually weakened the coupling between the electrode and the device.[47] Zhao et al.[48] observed improvement of the transport property of AGNR with the 5-8-5 (pentagon-octagon-pentagon) double vacancy defect, while ZGNR with 5-8-5 defect was reported to be unfavorable for electronic transport.

Recently the technological progress in highly focused and energetic electron and ion beam irradiation technique has made possible the controlled and selective generation of defects and monitoring structural reconstruction such as Stone-Wales defect vacancy defects, disorder etc in carbon based nano-structures.[32,49] Specifically divacancy (DV) defects with removal of two carbon atoms followed by a structural reconstruction is one of the most abundant defects in carbon based materials and thermodynamically more favorable than single vacancy defect.[45,50-51] Among the various possible configurations of DV defect such as 5-8-5 (two pentagons and one octagon), 555-777(three pentagons and three heptagons four pentagons), 5555-6-7777 (one hexagon and four heptagons), the 555-777 DV defect configuration is found to be the most stable one in GNR as reported *via ab-initio* simulations.[32,50,52]

In the present work, we investigated the modification of electronic structure of AGNR due to the introduction of DV 555-777 defect using state-of-the art density functional approach. We observed that there is a shifting of Fermi level towards the lower energy value, which is a signature of *p*-type doping. This result has motivated us to design and calculate the transport properties of AGNR based two-terminal device with one DV defected electrode and one N doped electrode. An asymmetric distribution of the electrostatic potential similar to conventional *p-n* junction device was observed across the scattering region. Our theoretically modeled device exhibits diode like property with high rectifying efficiency and also NDR phenomena with large peak-to-valley ratio has been observed.

## 2. Theoretical methodology

### 2.1 Model Structure

There exists three distinct groups of P-AGNR (where P is the number of dimmer lines across the ribbon width) *viz*. P=3n-1, 3n, 3n+1, with n integer, and we have considered mainly 8-AGNR, 9-AGNR and 10-AGNR in this study. For electronic structure calculation we took (1×1×4) supercell of the AGNRs as illustrated for pure 9-AGNR in Fig. 1a. 555-777 DV defect was introduced in the (1×1×4) supercell of the AGNRs, by removing two carbon atoms and rotating the bonds as required and the resulting structures were geometrically optimized. Fig. 1b shows the optimized structure of 555-777 DV defected 9-AGNR. All the edge carbon atoms of both pure and the defected AGNRs have been passivated by Hydrogen atoms. Spurious interactions between the periodic images are minimized in our theoretical models by considering a vacuum of greater than 15 Å along X and Y direction.

The model structure of the two terminal devices considered for transport calculation is shown in Fig. 1c. The system is divided into three parts, namely left-electrode, right-electrode and the scattering region. The shaded areas in Fig. 1c represent the semi-infinite electrodes. Right electrode was modeled by 555-777 DV defected AGNR, whereas a (1×1×4) AGNR supercell, doped with two N atoms was considered for modeling the left electrode. The scattering region was constructed by directly connecting the N doped AGNR to the defected AGNR. The scattering region created in this way is sandwiched between the two semi-infinite electrodes as shown in Fig. 1c.

## 2.2 Computational details

Geometry relaxation and the electronic structure calculations were performed using density functional theory (DFT) based code Vienna *Ab Initio* Simulation Package (VASP).[53] The exchange-correlation part was approximated by generalized gradient approximation (GGA) of Perdew, Burke, and Ernzerhof (PBE).[54] Projector augmented wave (PAW)[55] method was employed for describing the electron-ion interactions for the elemental constituents C, N, H. The plane wave basis cut off was 500 eV for all the calculations performed in this work. The Hellman−Feynman forces among the constituent atoms were minimized with the tolerance of 0.005 eV/Å, whereas the geometries of the two-terminal devices were optimized with a force tolerance of 0.03 eV/Å. The one dimensional Brillouin Zone (BZ) of the pure and defected AGNRs was sampled using Monkhorst-Pack methodology.[56] $1\times1\times5$ and $1\times1\times21$ k-mesh (periodic direction of the ribbon along the Z axis) was used during geometry optimization and Density of state calculations respectively.

Electronic transport properties of the two-terminal devices were computed using NEGF-DFT (Non-equilibrium Green's function method combined with DFT) technique as implemented in TranSIESTA code.[57] The optimized geometries of the two terminal devices, obtained from VASP code, were used in the transport calculations. As the size of the two terminal devices are very large, single-zeta (SZ) basis set has been used and the real space grid cutoff was set to 150 Ry in our transport calculations, in order to overcome the computational burden. In fact, SZ basis set yields reasonably good results for carbon based systems and has been used in several previously published reports.[58-60] For further justification of using SZ basis set, we performed some test calculations on defected 8-AGNR and N doped 8-AGNR with SZ basis (150 Ry mesh cutoff) as well as with double-zeta polarized (DZP) basis set (350 Ry mesh cutoff). The resulting band structures of SZ and DZP calculations were found to be nearly identical (see Fig. A1 and Fig. A2 of Appendix). Norm conserving Toullier-Martins pseudopotential[61] and PBE exchange-correlation functional was used during the calculation of transport properties. Self consistent calculation was carried out to obtain the current-voltage characteristics for the two-terminal devices with bias voltage ranging from -1.5 V to 1.5 V in steps of 0.1 V. The current through the scattering region at a finite bias ($V_b$) is calculated by integrating the transmission function at that bias within the bias energy window -e$V_b$/2 to +e$V_b$/2 using the Landauer-Buttikeformula[62]:

$$I(V_b) = \frac{2e}{h} \int_{-\infty}^{\infty} T(E, V_b)[f_l(E - \mu_l) - f_r(E - \mu_r)]dE,$$

Where $f_{l(r)}$ is the Fermi-Dirac distribution function for the left and right electrode and $\mu_{l(r)}$ is the electrochemical potential of the left (right) electrode such that $\mu_{l(r)} = E_f \pm e\frac{V_b}{2}$, with $E_f$ being the equilibrium Fermi energy of the system which was set to zero.

## 3. Results and Discussion

### 3.1 Defect induced modification in the electronic structure

First we discuss the modification of electronic structure of three different types of AGNR (*viz.* 8, 9, 10-AGNR) by DV 555-777 defect. The band structure and corresponding density of states (DOS) of pure and defected AGNRs are shown in Fig. 2. As revealed from Fig. 2a, 2c and 2e pristine AGNRs are semiconducting with energy band gap ($\Delta$) satisfying: $\Delta_{10\text{-AGNR}} > \Delta_{9\text{-AGNR}} > \Delta_{8\text{-AGNR}}$ and these results are in good agreement with the previous published theoretical report[22]. The reduction of structural symmetry due to the introduction of 555-777 DV defect in 8-AGNR results in electron-hole asymmetry as clearly observed from the band structure and DOS shown in Fig. 2b. The absence of electron-hole symmetry shifts the Fermi level of the defected 8-AGNR towards the lower energy value as compared with pure AGNR which is similar to that observed in case of *p*-type doping. As a result of 555-777 DV defect the highest $\pi$ and lowest $\pi^*$ band (indicated by two red bands in Fig. 2b) of pure 8-AGNR undergoes a shift towards the higher energy value with slight modification in their shapes and the energy gap between the above mentioned bands at the $\Gamma$ point increases by 0.12 eV as compared with the pure one. It is also evident from Fig. 2b that a localized defect state of band width 0.13 eV (green band) is introduced near the Fermi level and a corresponding sharp peak is observed at an energy 0.12 eV below $E_f$ in the DOS of defected 8-AGNR (right panel of Fig. 2b). The electronic structure of 9-AGNR and 10-AGNR get modified in a similar way due to the incorporation of 555-777 DV defect. The $\Gamma$ point energy gap between the highest $\pi$ and lowest $\pi^*$ band of pure 9-AGNR decreases by an amount of energy of 0.12 eV due to defect. For 9-AGNR the defect state (the green band in Fig. 2d) near $E_f$ is very localized having band width of only 0.06 eV and this localized band yields a peak in the DOS of defected 9-AGNR at -0.06 eV. However the defect induced change in the $\Gamma$ point energy gap, between the $\pi$ and $\pi^*$ band of 10-AGNR is very small, as is clearly evident from Fig. 2e and Fig. 2f. The band structure plot of defected 10-AGNR also shows that two localized bands crosses the Fermi level because of which there is a peak at $E_f$ in the DOS of defected 10-AGNR (Fig. 2f). Calculations performed with AGNR of larger width (11-AGNR, 12-AGNR, and 13-AGNR) also gave similar trends. So the main point to be emphasized here is that the introduction of 555-777 DV in three different classes of AGNR namely 3n-1, 3n, 3n+1 results in a shifting of the Fermi-level towards the lower energy value along with a upward shifting of the highest $\pi$ and lowest $\pi^*$ bands with a slight change in their shapes.

Now before going to discuss the transport properties of the device shown in Fig. 1c it is worth mentioning the electronic properties of N doped AGNR in brief. As N doping is preferred at the edge[63] we have substituted two edge C atoms with two N atoms for modeling the N-doped electrode. N doping in the AGNR brings additional electrons in the system and the Fermi level moves towards the higher energy value. Band structure and DOS plot of N doped AGNR of different width is given in the Fig. A3 of Appendix.

**3.2 Potential distribution across the scattering region**

In order to gain an insight into the internal polarization of the scattering region, we have analyzed the equilibrium electrostatic potential across the junction of defected and N doped electrode. The zero bias potential profile for the 9-AGNR junction is shown in Fig. 3. An asymmetric distribution of the potential is clearly observed from the 2-D potential plot (Fig. 3a). The color codes indicate that the N doped region is at a lower potential as compared with the defected region. In Fig. 3b we have plotted the potential, averaged along the X and Y direction, showing the oscillating behavior as indicated by the blue line. The averaged trend of the oscillating potential, denoted by the red line clearly shows the formation of a barrier across the junction, analogous to the case of conventional *p-n* junction. Therefore a natural rectifying character of the two-terminal device (shown in Fig. 1c) is expected. Equilibrium potential across the scattering region for the 8-AGNR and 10-AGNR devices shows similar trend (See Fig. A4 and A5 of Appendix).

### 3.3 Current-voltage characteristics

Now we discuss the transport properties of the AGNR based two-terminal devices. The current-voltage (I-V) characteristics of all the three AGNRs (8-AGNR, 9-AGNR, 10-AGNR) based devices are shown in Fig. 4. The asymmetric behavior of the current at positive and negative bias for all the devices, considered in our calculation, is clearly evident from the I-V curve shown in Fig. 4. This indicates the rectifying nature of the devices and will be discussed later. All the devices show the NDR phenomena besides their rectifying behavior, as clearly revealed from their I-V characteristics. For the 8-AGNR device under the application of positive bias the current increases and reaches to a value of 13.05 μA at 0.6 V and thereafter at 0.7 V current drops down to 8.35 μA as observed from Fig. 4a. So the first NDR phenomenon occurs in the bias range [0.6, 0.7] V and the corresponding peak-to-valley ratio (PVR) is 1.56, where PVR is defined as the ratio of $I_{max}$ to $I_{min}$ within a certain range of bias voltage. The second and the third NDR phenomena is observed in the bias ranges [0.8, 1.0] V and [1.3, 1.4] V and the corresponding PVR values are 2.87 and 1.03 respectively. For negative bias voltage the current increases almost monotonically in the range [0,-0.9] V, apart from the drop of current from 0.71 μA (at -0.1 V) to 0.002 μA (at -0.2 V). Also another noticeable NDR phenomenon with a PVR value of 23.68 is observed in the bias range [-0.8, -1] V. In case of 9-AGNR based device, in the positive bias region current increases monotonically in the bias range [0, 0.5] V and reaches a value of 15.13 μA at 0.5 V. With further increase in the positive bias, the current starts decreasing and drops to 0.0037 μA at 1.0 V (shown in Fig. 4b). So a strong NDR phenomena with a very high PVR value of 4089.18 occurs in [0.5, 1.0] V bias range. As clearly observed from Fig. 4b there are some peaks and valleys in the negative bias region [0, -0.7] V but the value of the current is quite small in this region. However there is a peak observed at -0.9 V where the current reaches a value of 3.68 μA and with further increase in negative bias the current starts decreasing and NDR phenomenon observed with a PVR of 42.66 in the bias range [-0.9, -1.3] V.

For the 10-AGNR device, NDR phenomena occurs two times in the positive bias region, in the range of [0.1, 0.6] V and the other in the range [1.3, 1.4] V (seen from Fig. 4c) with respective PVR of 12828.29 and 2.15. In the negative bias region NDR appears four times in the ranges [-0.1, -0.2] V, [-0.3, -0.4] V, [-0.6, -0.7] V, [-0.9, -0.1.1] V with the respective PVR of 1.08, 382.10, 33.48 and 4.79.

Now we focus our attention on the rectifying efficiency of the two-terminal devices. The rectification ratio (RR) is defined as RR(V) = |I(V)/I(-V)| for positive (forward) rectification and RR(V) = -|I(-V)/I(V)| for negative (reversed) rectification. The rectification ratios for all the three AGNRs (8-AGNR, 9AGNR, 10-AGNR) based devices are given is Table 1. The 8-AGNR based device shows positive rectification for all the biases except at 0.1 V and it shows a highest rectifying efficiency of 1213.25 at 0.2 V. The 9-AGNR based device shows rectification in the forward direction in the bias region [0.1, 0.7] V as well as in [1.1, 1.5] V bias region. However it shows reversed rectification in region [0.8, 1.0] V. It should be noted that the current drops to a very small value in the positive bias range [0.8, 1.0] V (NDR phenomena with a very large PVR occurs in the range of [0.5, 1.0] V as already discussed), whereas in the negative bias range [-0.8, -1.0] V the value of the current is quite high. So a negative rectification is observed for the 9-AGNR based device in the above mentioned region. The highest rectification ratio of the 9-AGNR based device is 7272.57 at 0.5 V. For the 10-AGNR based rectification in the reversed direction with very high RR is observed for a wide range of bias [0.4, 1.2] V. The highest RR value of -12391.03 is observed at 0.6V for the 10-AGNR device. Here the point to be mentioned is that rectifiers with length, scaled down to a few nanometer, can show negative rectification as reported in several theoretical studies.[40,44,64] The rectifying efficiency of all the three AGNRs based devices are quite high, making them potential candidate for nano-electronic device applications.

### 3.4 Interpretation *via* transmission function

In order to elucidate the nonlinearity and asymmetry observed in the I-V characteristics, we have plotted (in Fig. 5) the transmission functions (T(E)) along with the DOS of the left and right electrodes, at six different bias voltages for the 9-AGNR device. At zero bias there are two peaks in T(E) around the Fermi level as observed from Fig. 5a; but since the bias window is zero the current is also zero. It should be noted that under the application of positive bias $V_b$, the chemical potential of the left (right) electrode increases (decreases) by $eV_b/2$. So with applied positive bias, the energy states of the left and right electrode shift respectively towards the right and left direction with respect to the equilibrium Fermi level. Consequently more and more resonant states of the two electrodes appear in the bias window, on slowly increasing the positive bias (in the region [0, 0.5] V). This results in a gradual growth of the T(E) around the Fermi level. Particularly at 0.5 V there is a very good matching of the left electrode DOS (LDOS) and right electrode DOS (RDOS) within the bias window, as indicated by the shaded region in Fig. 5b. This well matched energy states of the two electrodes leads to the formation of a pronounced peak in the transmission function at the Fermi level (Fig. 5b) which results in a very high value of current at that bias (see Fig. 4b). With further increase in the positive bias, the energy states of the left (right) electrode move further to the higher (lower) energy value resulting in mismatch of the earlier well matched states. So with the gradual increase in the bias in the region [0.5, 1.0] V, the value of T(E) within the bias window gradually decreases. It can be noticed from Fig. 5c, that there is no significant matching between LDOS and RDOS at 1.0 V, within the bias window and accordingly the value of T(E) is found to be negligibly small. As a result, the current drops to very small value, thereby leading to a NDR phenomenon with a very high PVR in the region [0.5, 1.0] V (discussed in the previous section). With the bias voltage increased further some matched states of the two electrodes comes into the bias window, leading to the development of two peaks in T(E) one at -0.55 eV and another at 0.55 eV. Hence a monotonic increasing behavior of current has been observed in the bias range [1.0, 1.5] V (see Fig. 4b). At 1.5 V bias, the good matching between the LDOS and RDOS are shown in Fig. 5d (indicated by the shading) and the two pronounced transmission peaks are also clearly observed at -0.55 eV and 0.55 eV. On the other hand if the bias voltage is applied in the reversed direction, the chemical potential of the left (right) electrode decreases (increases) by $eV_b/2$ ($V_b$ is the applied negative bias) and consequently the energy states of the left (right) electrode moves towards left (right) with respect to the equilibrium Fermi level. Hence the matching between the states of the two electrodes turns out to be very poor. For example at -0.5 V there is no matching between the LDOS and RDOS (Fig. 5e) and the resulting T(E) yields negligibly small value within the bias window. The small value of T(E) and concomitant low current in the negative bias as compared with the high current in positive bias (Fig. 4b) explains the high rectifying efficiency of the device. However, at -0.9 V there is a noticeable peak in the current (Fig. 4b) originating from the synchronized states inside the bias window (as indicated by the shaded region in Fig. 5f). Thus the 'matching' and 'mismatching' of the electrode energy states within the bias window qualitatively dictates the development and suppression of the transmission function. This bias dependent transmission function (within the bias window) provides a clear picture of the nonlinear and asymmetric behavior observed in the I-V characteristics. Similar conclusion can be derived for the 8-AGNR and 10-AGNR based devices, whose bias dependent transmission functions are shown in Fig. A6 and A7 of Appendix.

## 4. Conclusions

In summary, we have studied the 555-777 DV defect induced modification in the electronic structure of AGNRs and the transport properties of AGNR based two-terminal devices. It has been observed that the DV defect create electron-hole asymmetry which moves the highest $\pi$ and lowest $\pi^*$ band of AGNRs to the higher energy value. This causes shifting of Fermi level towards the lower energy value, similar to the case of *p*-type doping. Electrostatic potential distribution across the scattering region, constructed by N doped AGNR and 555-777 DV defected AGNR shows asymmetric feature, similar to the case of conventional *p-n* junction. NDR phenomenon has been found for all the three AGNRs (8-AGNR, 9-AGNR, 10-AGNR) based devices in both positive and negative bias. Mostly positive rectification for 8-AGNR and 9-AGNR devices with highest rectification ratio of 1213.25 and 7272.57 respectively have been estimated. Whereas the 10-AGNR based device shows negative rectification for a wide range of bias and the highest calculated RR is -12391.03 at 0.6V. The modification of transmission function with applied bias, based on the 'matching' and 'mismatching' of the energy states of left and right electrodes plays, the key role in determining the observed transport properties. Hence the DV defected AGNR based two-terminal device can be effectively utilized to design nano-electronic devices.

**Acknowledgement**

SC is financially supported by a CSIR fellowship 09/080(0787)/2011-EMR-I. RT thanks SERB for the financial support (Grant no: SB/FTP/PS028/2013). GPD acknowledges the support from IBIQUS project from the Dept. of Atomic Energy, Govt. of India (DAE). The authors thank Dr. Chiranjib Majumder (BARC) and Dr. Rajib Batabyal (IACS) for many helpful discussions.

**Appendix**

Modification of the Electronic structures of AGNRs due to N doping is shown. 2-D potential (at zero bias) distribution across the scattering region, left electrode density of states, right electrode density of states and transmission function of 8-AGNR and 10-AGNR based devices are shown in detail.

**Table 1. Rectification ratios of AGNR based devices at different bias.**

| Bias (V) | Rectification Ratio | | |
|---|---|---|---|
| | 8-AGNR | 9-AGNR | 10-AGNR |
| 0.1 | -3.85 | 10.59 | 5.09 |
| 0.2 | 1213.25 | 4129.26 | 4.93 |
| 0.3 | 239.14 | 14.65 | -2.09 |
| 0.4 | 6.64 | 31.64 | -6.72 |
| 0.5 | 8.37 | 7272.57 | -1154.41 |
| 0.6 | 5.67 | 5.79 | -12391.03 |
| 0.7 | 2.64 | 18.45 | -353.16 |
| 0.8 | 1.95 | -4.24 | -360.23 |
| 0.9 | 29.94 | -142.66 | -1400.72 |
| 1 | 16.14 | -432.89 | -698.23 |
| 1.1 | 9.72 | 1.69 | -71.34 |
| 1.2 | 31.16 | 15.64 | -290.49 |
| 1.3 | 16.40 | 143.71 | 2.38 |
| 1.4 | 8.65 | 4.76 | -1.25 |
| 1.5 | 25.77 | 4.66 | 1.33 |

**Figures and Captions**

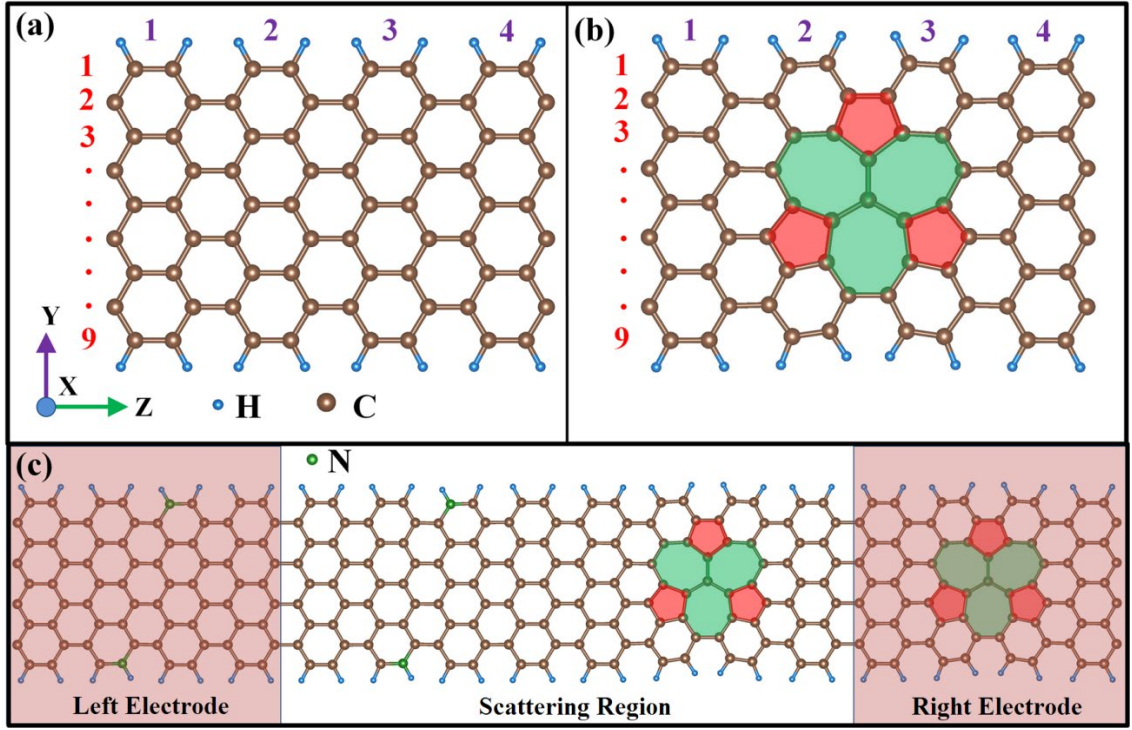

**Fig. 1** (a) and (b) demonstrate the optimized structure of (1×1×4) supercell of 9-AGNR and 555-777 divacancy defective 9-AGNR, (c) shows the model structure of the two-terminal device constructed by N doped AGNR connected with 555-777 divacancy defected AGNR. The shaded region indicate the left and right electrode. The blue, brown and green sphere H, C and N atoms respectively. The Z direction represents the transport direction.

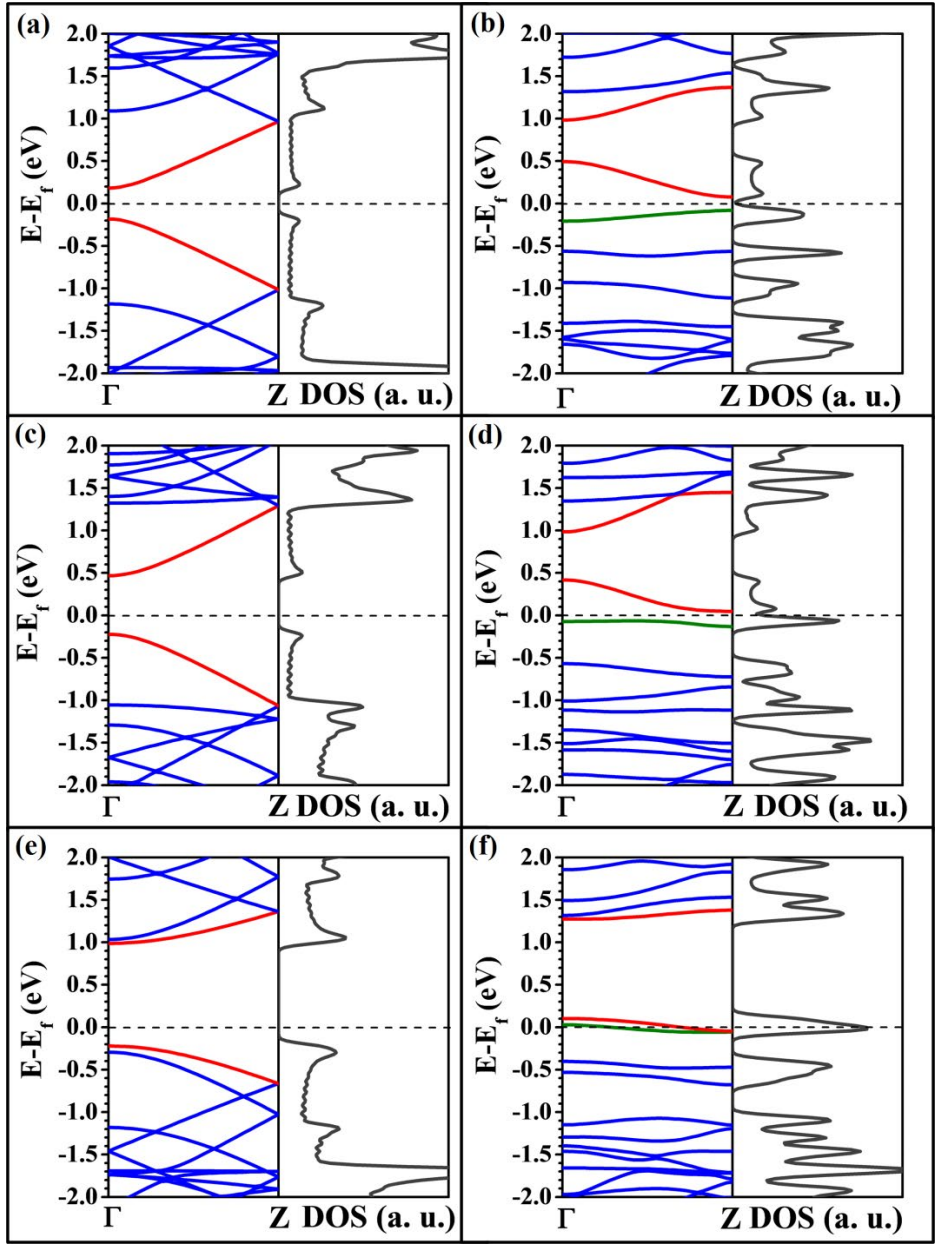

**Fig. 2** (a), (c) and (e) show the band structure and density of states of pure 8-AGNR, 9-AGNR and 10-AGNR, (b), (d) and (f) show the band structure and density of states of 555-777 DV defected 8-AGNR, 9-AGNR and 10-AGNR respectively. The red bands represent the π and π* bands of pure AGNRs and the corresponding bands of the DV defected AGNRs. The green bands represent the defect levels. The Fermi level is set to zero.

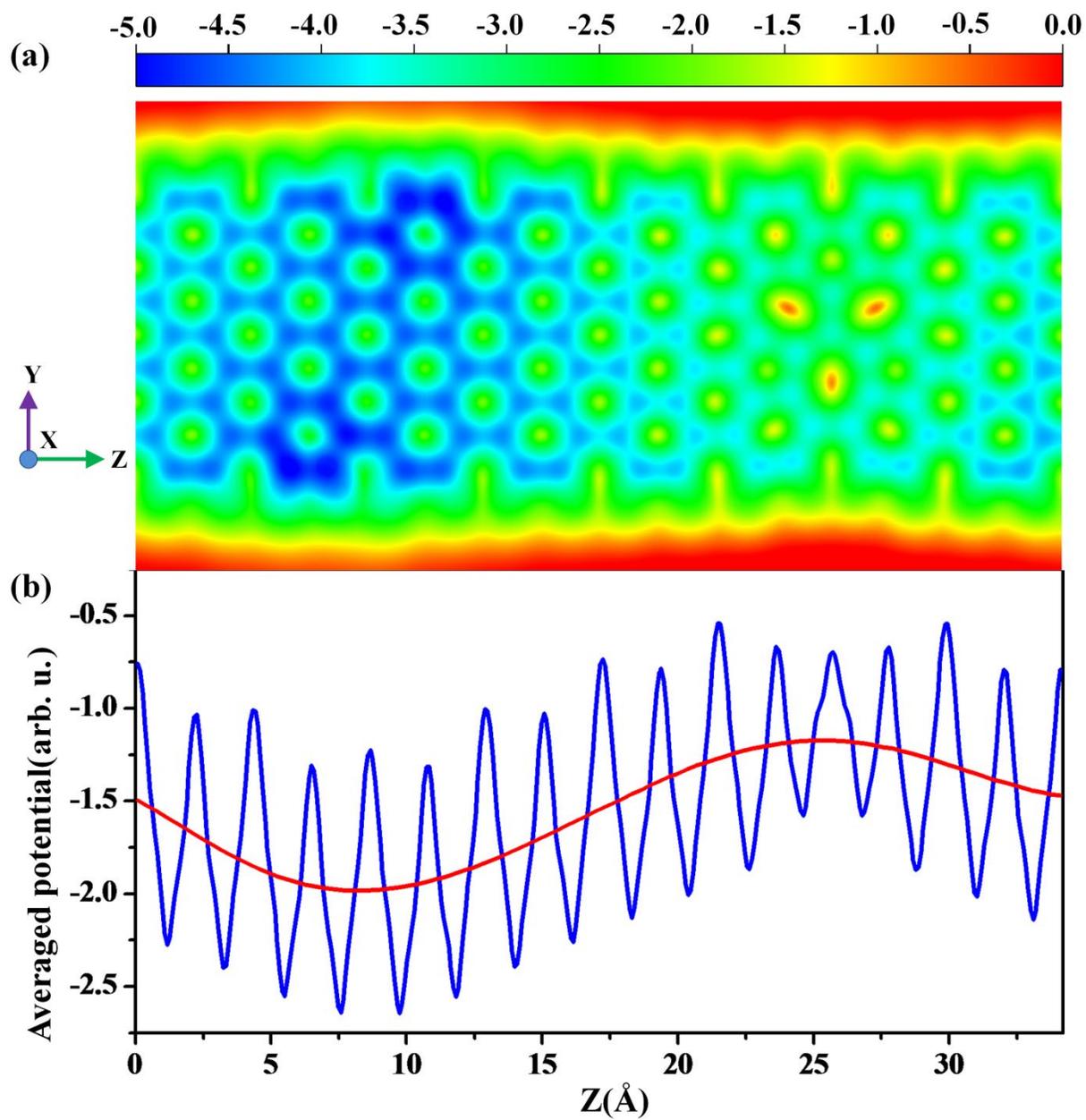

**Fig. 3** (a) 2-D potential distribution of the 555-777 DV defected and N doped 9-AGNRs junction at equilibrium, (b) Potential averaged along X and Y direction. The red line is to indicate the average trend of the oscillating potential.

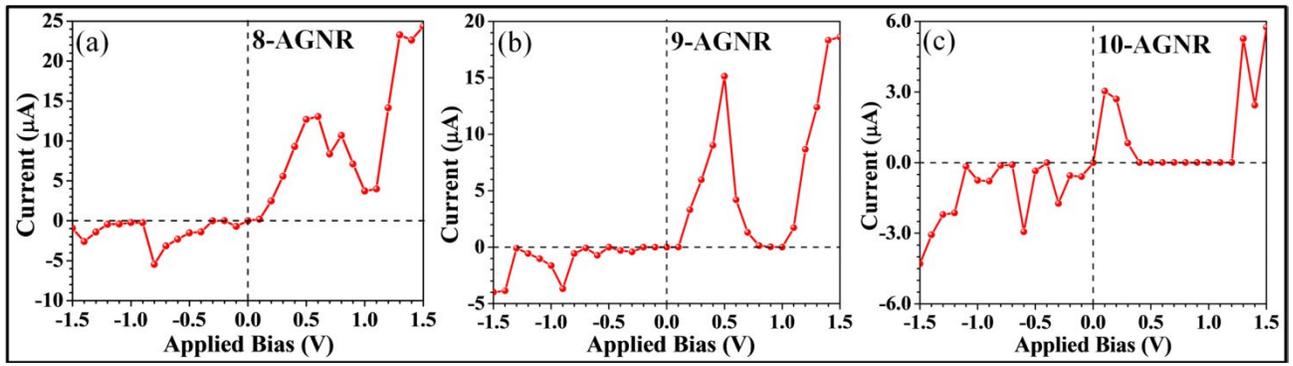

**Fig. 4** I-V characteristics of (a) 8-AGNR, (b) 9-AGNR and (c) 10-AGNR based two- terminal devices.

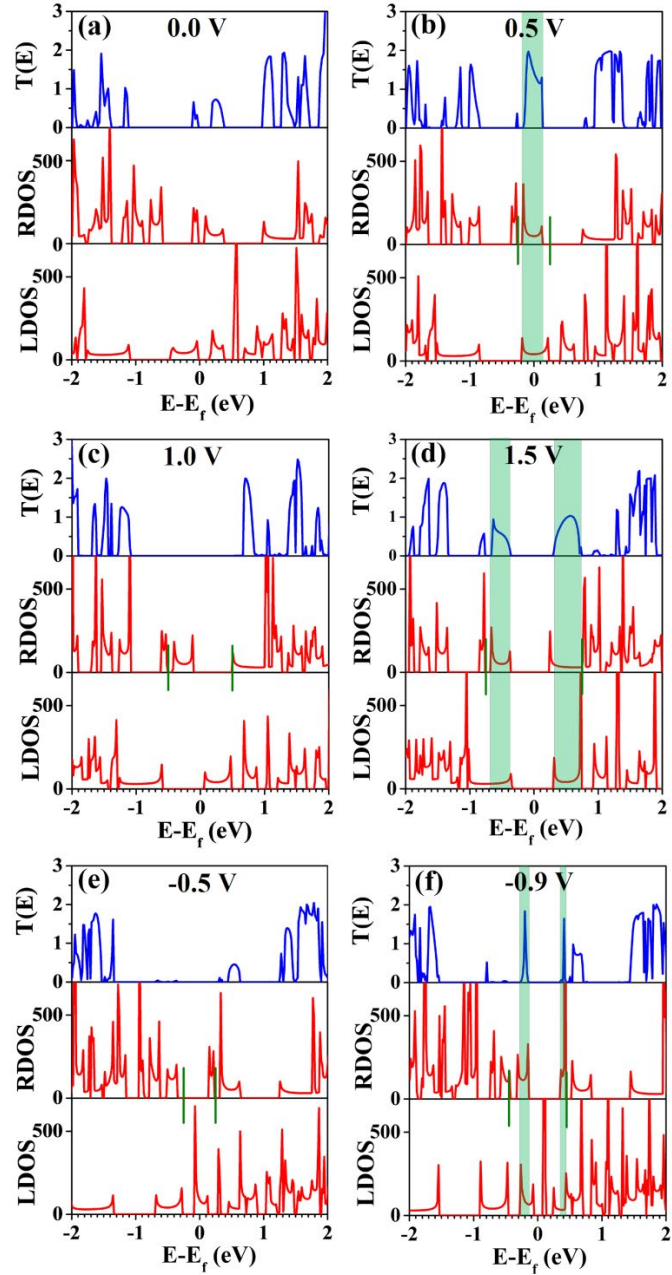

**Fig. 5** Left electrode density of states (LDOS), right electrode density of states (RDOS), transmission function (T(E)) at bias (a) 0 V, (b) 0.5 V, (c) 1.0 V, (d) 1.5 V, (e) -0.5 V and (f) -0.9 V. The two green vertical lines indicate the bias window at each bias voltage. The shaded regions indicate the matching of LDOS and RDOS within the bias window at each bias voltage.

**APPENDIX**

In order to ensure continuity of the essential physics in this paper, we have included these figures in the APPENDIX.

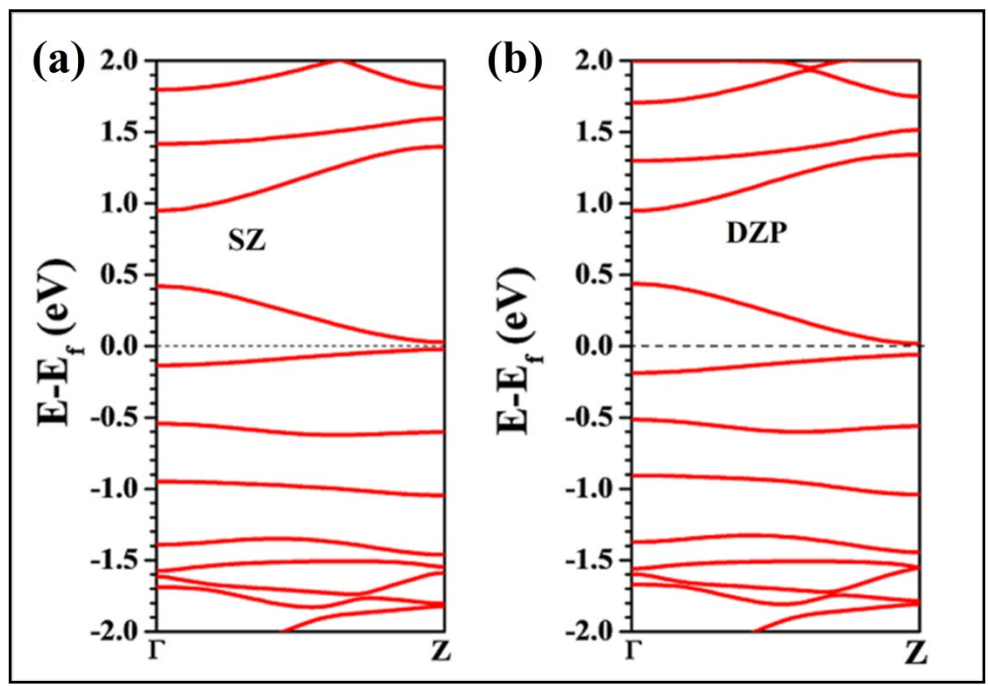

**Fig. A1** Band structure plots of 555-777 divacancy defected 8-AGNR. (a) shows the SIESTA band structure with SZ basis and 150 Ry mesh cutoff. (b) shows the SIESTA band structure with DZP basis and 350 Ry mesh cutoff.

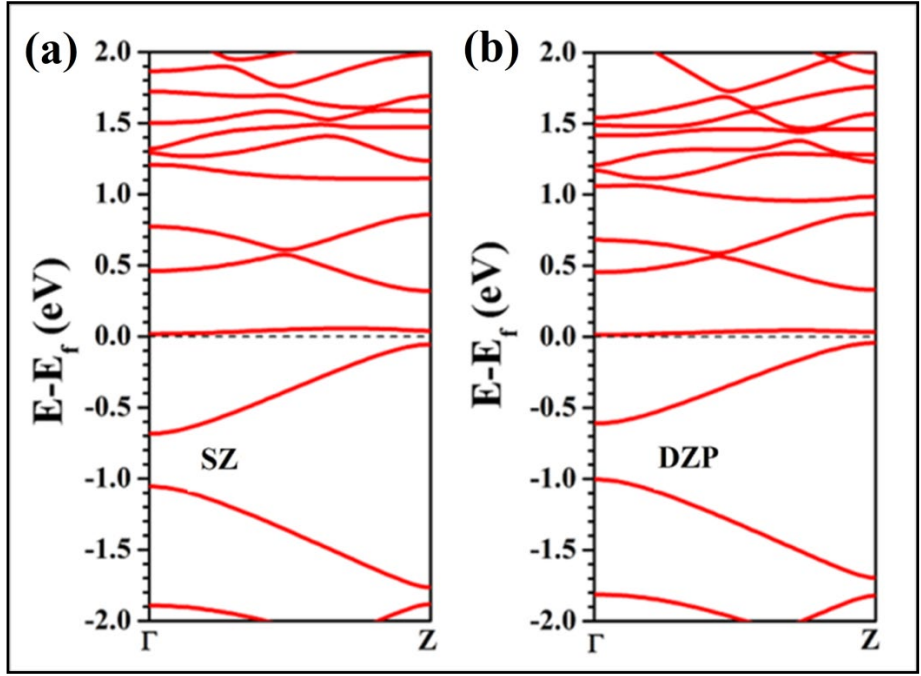

**Fig. A2** Band structure of *N* doped 8-AGNR. (a) shows the SIESTA band structure with SZ basis and 150 Ry mesh cutoff. (b) shows the SIESTA band structure with DZP basis and 350 Ry mesh cutoff.

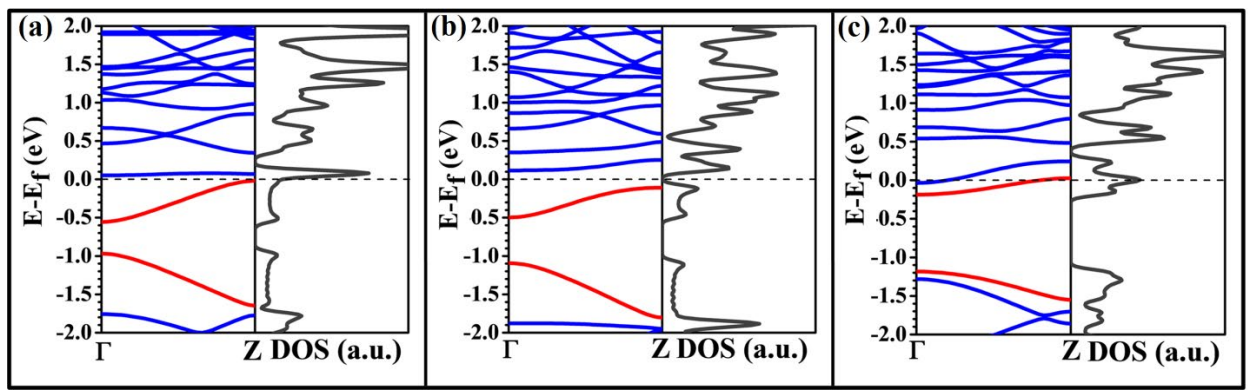

**Fig. A3** (a), (b) and (c) show the band structure and density of states of N doped 8-AGNR, 9-AGNR and 10-AGNR respectively. The red bands indicates the modified and downward shifted π and π* bands of the pure AGNRs due to N doping.

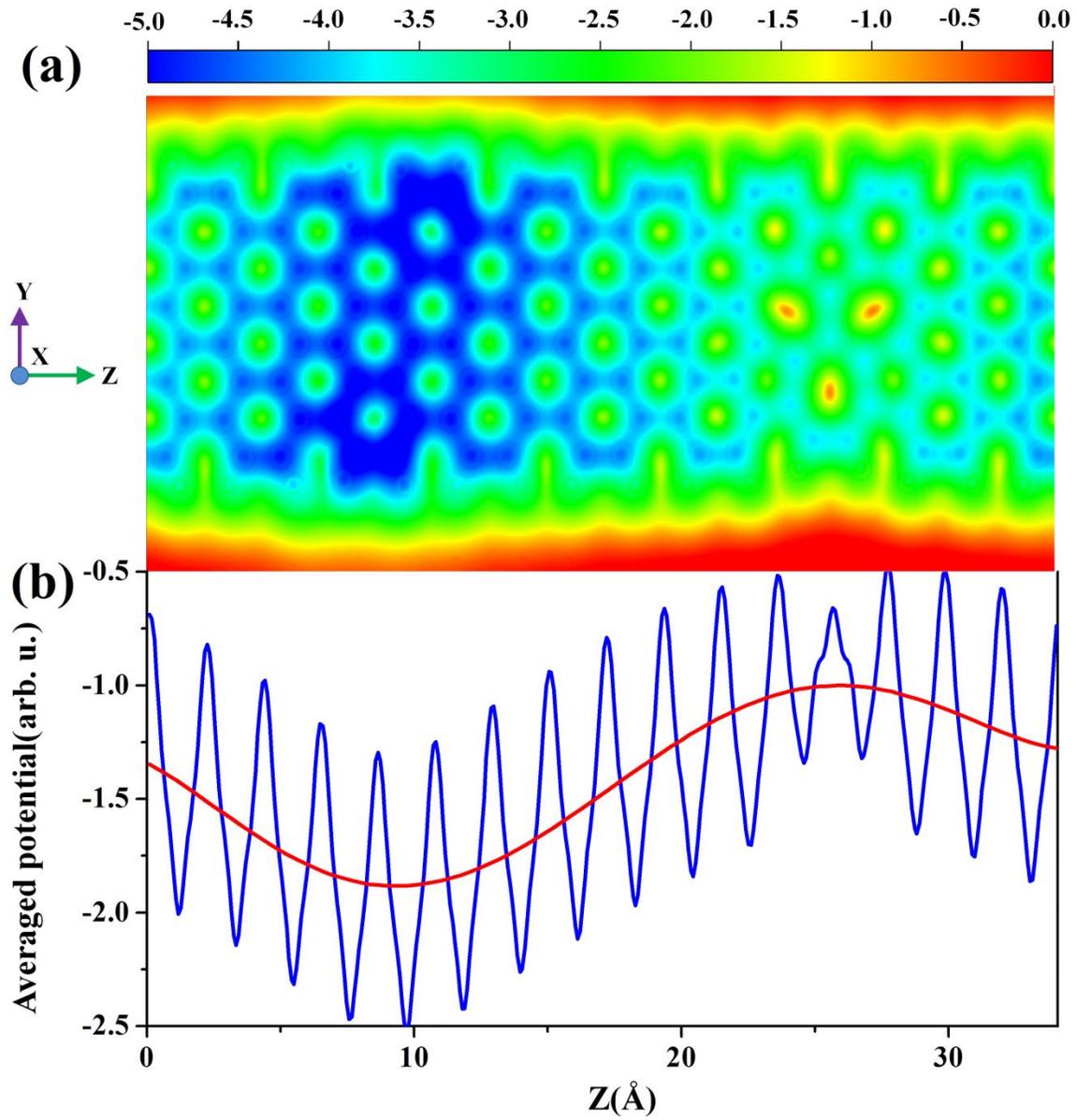

**Fig. A4** (a) 2-D potential (at zero bias) distribution across the scattering region of 8-AGNR based device, (b) Potential averaged along X and Y direction. The red line is to indicate the averaged trend of the oscillating potential.

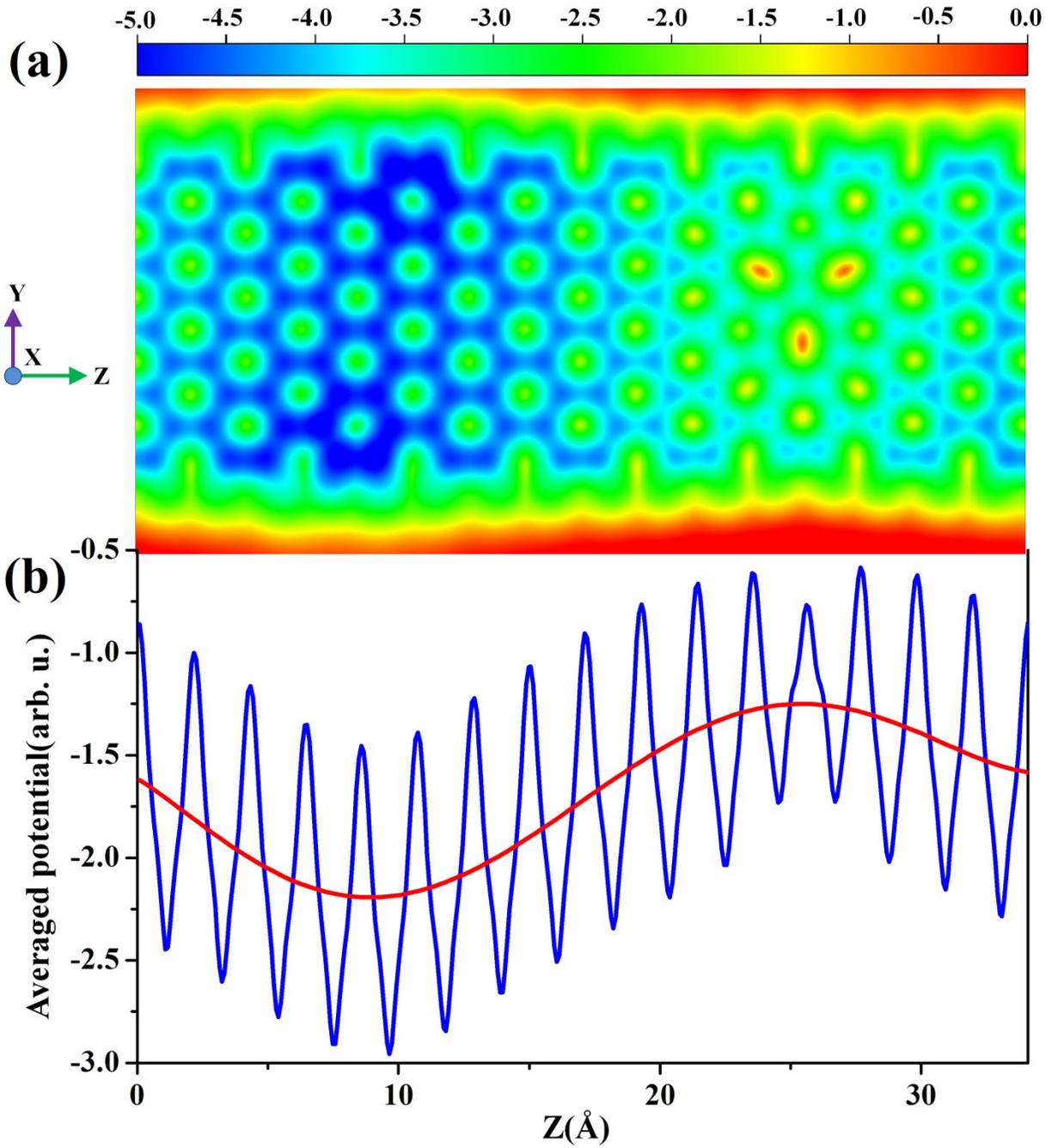

**Fig. A5** (a) 2-D potential (at zero bias) distribution across the scattering region of 10-AGNR based device, (b) Potential averaged along X and Y direction. The red line is to indicate the averaged trend of the oscillating potential.

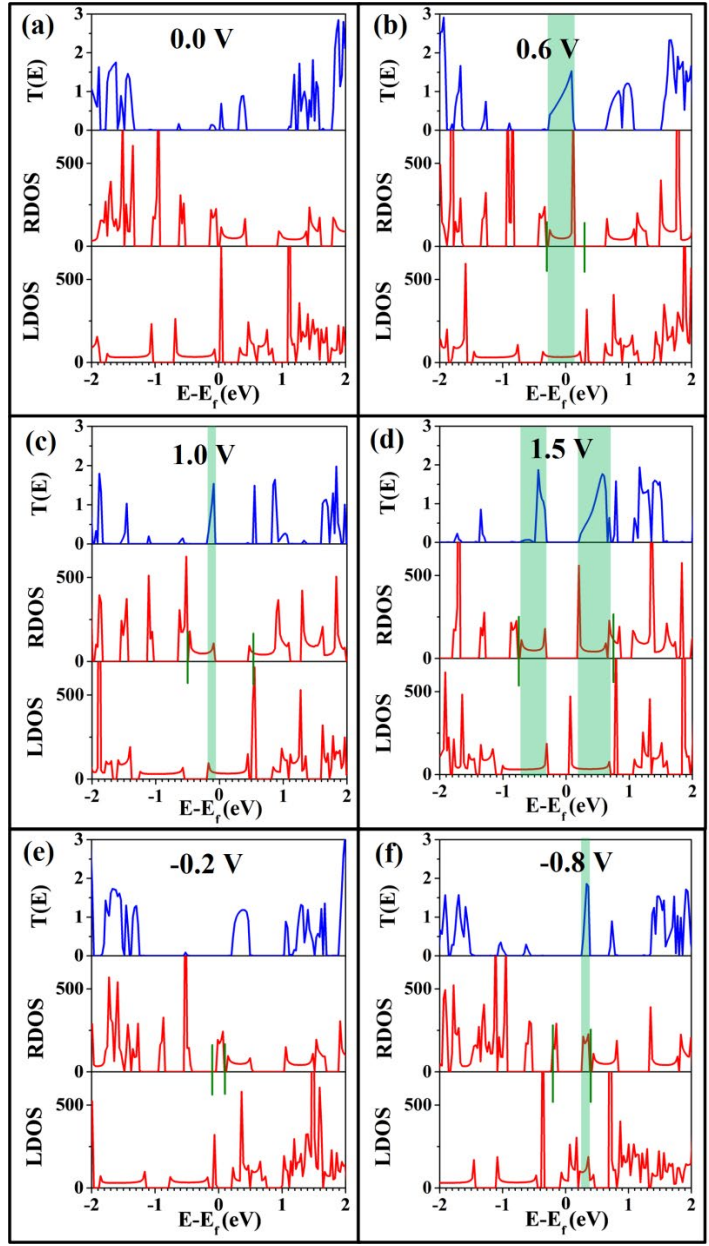

**Fig. A6** Left electrode density of states (LDOS), right electrode density of states (RDOS) and transmission function (T(E)) for the 8-AGNR based two-terminal device at bias (a) 0 V, (b) 0.6 V, (c) 1.0 V, (d) 1.5 V, (e) -0.2 V and (f) -0.8 V. The two green vertical lines indicate the bias window at each bias voltage. The shaded regions indicate the matching of LDOS and RDOS within the bias window at each bias voltage.

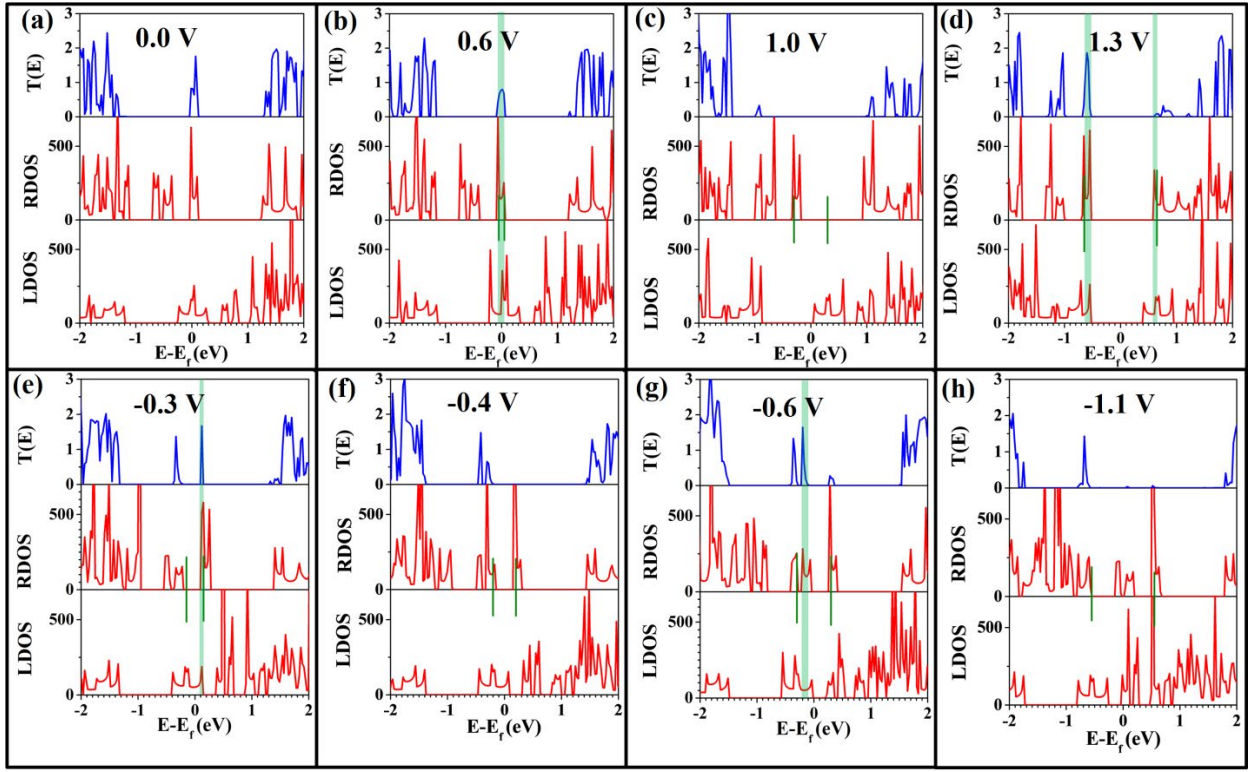

**Fig. A7** Left electrode density of states (LDOS), right electrode density of states (RDOS) and transmission function (T(E)) for the 10-AGNR based two-terminal device at bias (a) 0 V, (b) 0.6 V, (c) 1.0 V, (d) 1.3 V, (e) -0.3 V, (f) -0.4 V, (g) -0.6 V and (h) -1.1 V. The two green vertical lines indicate the bias window at each bias voltage. The shaded regions indicate the matching of LDOS and RDOS within the bias window at each bias voltage.